\documentclass[prl,twocolumn,superscriptaddress]{revtex4}

\usepackage{graphicx}
\usepackage{amssymb}
\usepackage{times}
\usepackage{amsmath}

\begin{document}

\title{Quantum Cryptography Without Switching}

\author{Christian Weedbrook} \affiliation{Quantum Optics Group,
Department of Physics, Faculty of Science, Australian National
University, ACT 0200, Australia}
\author{Andrew M. Lance} \affiliation{Quantum Optics Group, Department
of Physics, Faculty of Science, Australian National University, ACT
0200, Australia}
\author{Warwick P. Bowen} \affiliation{Quantum Optics Group,
Department of Physics, Faculty of Science, Australian National
University, ACT 0200, Australia}
\author{Thomas Symul} \affiliation{Quantum Optics Group, Department of
Physics, Faculty of Science, Australian National University, ACT 0200,
Australia}
\author{Timothy C. Ralph} \affiliation{Department of Physics,
University of Queensland, St Lucia, Queensland 4072, Australia}
\author{Ping Koy Lam} \affiliation{Quantum Optics Group, Department of
Physics, Faculty of Science, Australian National University, ACT 0200,
Australia}

\date{\today}

\begin{abstract}
We propose a new coherent state quantum key distribution protocol that
eliminates the need to randomly switch between measurement bases.
This protocol provides significantly higher secret key rates with
increased bandwidths than previous schemes that only make single
quadrature measurements.  It also offers the further advantage of
simplicity compared to all previous protocols which, to date, have
relied on switching.
\end{abstract}

\pacs{03.67.Dd, 42.50.Dv, 89.70.+c}

\maketitle

Quantum cryptography is the science of sending secret messages via a
quantum channel.  It uses properties of quantum mechanics
\cite{Wiesner,BB84} to establish a secure key, a process known as
quantum key distribution (QKD) \cite{QKD}.  This key can then be used
to send encrypted information.  In a generic QKD protocol, a sender
(Alice) prepares quantum states which are sent to a receiver (Bob)
through a potentially noisy channel.  Alice and Bob agree on a set of
non-commuting bases to measure the states with.  Using various
reconciliation \cite{Cerf} and privacy amplification \cite{Brassard}
procedures, the results of measurements in these bases are used to
construct a secret key, known only to Alice and Bob.  Switching
randomly between a pair of non-commuting measurement bases ensures
security: in a direct attack, an eavesdropper (Eve) will only choose
the correct basis half the time; alternatively, if Eve uses quantum
memory and performs her measurements after Bob declares his basis, she
is unable to manipulate what Bob measures.  It is commonly assumed
that randomly switching between measurement bases is crucial to the
success of QKD protocols.  In this Letter we show
that this is not the case, and in fact greater secret key rates can be
achieved by simultaneously measuring both bases.

The original QKD schemes in the discrete variable regime were based on
the transmission and measurement of random polarizations of single
photon states \cite{BB84}.  Other discrete variable QKD protocols have
been proposed \cite{Eke91} and experimentally demonstrated
\cite{Nai00} using Bell states.  However the bandwidth of such schemes
is experimentally limited by single photon generation and detection
techniques.  Consequently in the last few years there has been
considerable interest in the field of continuous variable quantum
cryptography \cite{Brau03}, which provides an alternative to the
discrete approach and promises higher key rates.  Continuous variable
QKD protocols have been proposed for squeezed and
Einstein-Podolsky-Rosen entangled states
\cite{Hil00,Cer01,Sil02,Rei00,Got01}.  However, these protocols
require significant quantum resources and are susceptible to
decoherence due to losses.  QKD protocols using
coherent states were proposed to overcome these limitations.
Originally such schemes were only secure for line losses less than
50\% or 3dB \cite{ral00}.  This apparent limitation was overcome using
the secret key distillation techniques of post-selection
\cite{Silberhorn&Ralph3dB} and reverse reconciliation
\cite{G&GReverse}.
\begin{figure}[!ht]
\begin{center}
\includegraphics[width=\columnwidth]{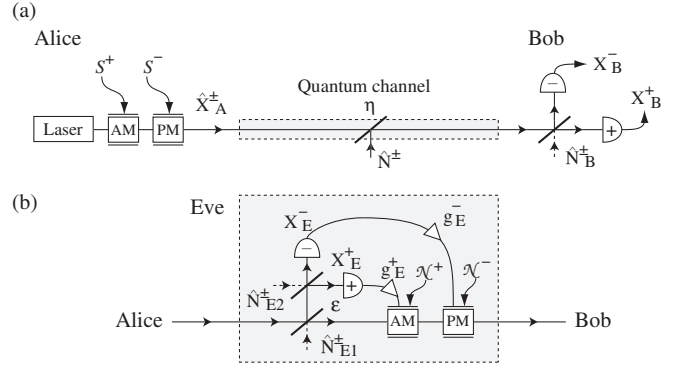}
\caption{(a) Schematic of the simultaneous quadrature measurement
protocol. $\mathcal{S}^{\pm}$: random Gaussian numbers, AM: amplitude
modulator, PM: phase modulator, $\hat{X}_{A}^{\pm}$: quadratures of
Alice's prepared state, $\eta$: channel transmission, $\hat{N}^{\pm}$:
channel noise, $\hat{X}_{B}^{\pm}$: observables that Bob measures and
$\hat{N}_{B}^{\pm}$: Bob's vacuum noise. (b) Schematic of a possible
feed forward attack for Eve. $\hat{X}_{E}^{\pm}$: observables that Eve
measures, $\hat{N}_{E1}^{\pm}$ and $\hat{N}_{E2}^{\pm}$: Eve's vacuum noises,
$g_{E}^{\pm}$: Eve's electronic gains and $\mathcal{N}^{\pm}$:
additional Gaussian noise.}\label{FigureSchematic}
\end{center}
\end{figure}

In general, security in discrete variable cryptography protocols is
ensured via random switching between measurement bases \cite{BB84} or
random switching of state manipulation \cite{Bos02}.  The random
switching between measurement bases can be achieved simply via a 50/50
beam splitter, where the selection of the measurement basis is chosen
through the random photon transmission and reflection statistics.  To
date all continuous variable cryptography protocols have also relied
on randomly switching between non-commuting bases.  In the continuous
variable regime, switching requires precise control of the phase of a
local oscillator beam, which is difficult to achieve in practice.
This local oscillator switching currently places a serious technical
limitation on the bandwidth of cryptography protocols.  In this
Letter, we introduce a new coherent state protocol that does not
require switching.  In this protocol, both bases are measured
simultaneously, utilizing the quantum channel more effectively and
achieving both higher secret key rates and bandwidths compared to
previous continuous variable QKD protocols.

The quantum states we consider in this letter can be described using
the field annihilation operator
$\hat{a}\!=\!(\hat{X}^{+}\!+i\!\hat{X}^{-})/2$, which is expressed in
terms of the amplitude $\hat{X}^{+}$ and phase $\hat{X}^{-}$
quadrature operators.  In this paper, we denote operators and real
numbers with and without (~$\hat{}$~), respectively, to avoid
confusion.  Without a loss of generality, the quadrature operators can
be expressed in terms of a steady state and fluctuating component as
$\hat{X}^{\pm}\!=\!\langle\hat{X}^{\pm}\rangle\!+\!\delta\hat{X}^{\pm}$,
which have variances of
$V^{\pm}\!=\!\langle(\delta\hat{X}^{\pm})^2\rangle$.
Figure~\ref{FigureSchematic}(a) shows a schematic of our protocol,
which we term the simultaneous quadrature measurement protocol.  Our
scheme is similar to the continuous variable coherent state quantum
cryptography protocols presented in \cite{ral00}.  The protocol
goes as follows: Alice draws two random real numbers $\mathcal{S}^{+}$
and $\mathcal{S}^{-}$ from Gaussian distributions with zero mean and a
variance of $V^{\pm}_{\mathcal{S}}$.  She then prepares a state by
displacing the amplitude and phase quadratures of a vacuum state by
$\mathcal{S}^{+}$ and $\mathcal{S}^{-}$, respectively.  The quadrature
operators of Alice's state are therefore given by $\hat{X}^{\pm}_{
A}\!=\!\mathcal{S}^{\pm}\!+\!\hat{X}^{\pm}_{v}$, where
$\hat{X}^{\pm}_{v}$ are the quadrature operators of the initial
vacuum state.  The resulting state has normalized quadrature variances
of $V^{\pm}_{A}=V^{\pm}_{\mathcal{S}}+1$.  Alice transmits this
state to Bob through a quantum channel with channel transmission
efficiency $\eta$, which couples in channel noise $\hat{N}^{\pm}$,
where the variances of the channel noise must obey the uncertainty
relation $V^{+}_{N} V^{-}_{N} \geq 1$.  Bob simultaneously measures
the amplitude and phase quadratures of the state using a 50/50 beam
splitter.  The quadrature variances of the state measured by Bob are
given by
\begin{equation}\label{Bobvar}
   V_{B}^{\pm}=\frac{1}{2}\Big(\eta
V_{A}^{\pm}+(1-\eta)V_{N}^{\pm}+1\Big)
\end{equation}
By using secret key distillations protocols {\cite{Cerf}} and
standard privacy amplification techniques {\cite{Brassard}},
Alice and Bob can then distill a common secret key.
\begin{figure}[!ht]
\begin{center}
\includegraphics[width=\columnwidth]{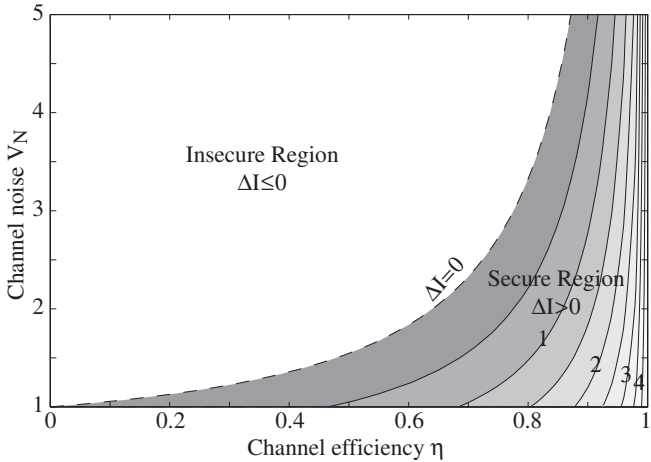}
\caption{Contour plot of the information rate for the simultaneous
quadrature measurement protocol as a function of channel
efficiency $\eta$ and channel noise $V_{N}$ in units of
(bits/symbol) for $V_{A}=100$.}\label{FigureContour}
\end{center}
\end{figure}
It is possible to analyze our protocol using either the
post-selection, or reverse reconciliation, secret key distillation
techniques \cite{G&GReverse,Silberhorn&Ralph3dB}.  However,
for simplicity, we limit our analysis to the Grosshans and Grangier
reverse reconciliation protocol \cite{G&GReverse}.  In this
protocol, Alice and Eve both try to infer Bob's measurement results.
Alice's inference can be characterized by a conditional variance which
is used to calculate the secret key rate.  Alice's conditional variance 
given Bob's measurement
can be written as $V^{\pm}_{A|B}=\min_{g^{\pm}_{A}}
\langle(\hat{X}^{\pm}_{B}-g^{\pm}_{
A}\hat{X}^{\pm}_{A})^{2}\rangle$, where the gain $g^{\pm}_{\rm A}$ is
optimized to give a minimum conditional variance of 
%
%
\begin{equation}\label{DefAliceCondVar}
V^{\pm}_{A|B}=V^{\pm}_{B}-{|{\langle
\hat{X}^{\pm}_{A}\hat{X}^{\pm}_{B}\rangle}|^{2}}/{V^{\pm}_{A}}
\end{equation}
To calculate a relation between Alice's and Eve's conditional
variances of Bob's measurement, $V_{E|B}^{\pm}$ and $V_{A|B}^{\pm}$,
we define the states that denote Alice's and Eve's inference of Bob's
measurement, expressed as:
$\hat{X}_{E|B}^{\pm}=\hat{X}_{B}^{\pm}-\alpha \hat{X}_{E}^{\pm}$ and
$\hat{X}_{A|B}^{\mp}=\hat{X}_{B}^{\mp}-\beta \hat{X}_{A}^{\mp}$ where
$\beta \hat{X}_{A}^{\pm}$ and $\alpha \hat{X}_{E}^{\pm}$ are Alice and
Eve's optimal estimates with the optimal gains, $\alpha$ and $\beta$
being real numbers.  Finding the commutator of these two equations,
and using the fact that different Hilbert spaces commute, we find that
$[\hat{X}_{E|B}^{+},\hat{X}_{A|B}^{-}]=[\hat{X}_{B}^{+},\hat{X}_{B}^{-}]=2i$
\cite{G&GReverse}.  This leads to the joint Heisenberg uncertainty
relation
\begin{equation}\label{VarUncertainty}
V_{E|B}^{\pm}V_{A|B}^{\mp}\geq1
\end{equation}
Therefore, there is a limit to what Alice and Eve can know
simultaneously about what Bob has measured.  From this inequality it
is possible to determine the maximum information Eve can obtain about
the state in terms of Alice's conditional variances $V_{A|B}^{\pm}$.

To minimize Alice's conditional variance for one of Bob's
measurements, Alice can prepare and send squeezed states, instead of
coherent states.  In this case, the quadrature variance of the states
prepared by Alice are given by $V^{\pm}_{A}=V^{\pm}_{{\rm
\mathcal{S}}}+V^{\pm}_{{\rm sqz}}$, where $V^{\pm}_{{\rm sqz}}$ denote
the quadrature variances of the squeezed state, and clearly $V_{{\rm
sqz}}^{\pm} \geq 1/V_{A}^{\mp}$.  Using Eqs.~(\ref{Bobvar}) and
(\ref{DefAliceCondVar}) we determine Alice's conditional variance to
be
\begin{eqnarray}\label{AlicesCondVar}
V^{\pm}_{A|B}&=&\frac{1}{2}\Big(\eta V^{\pm}_{\rm
sqz}+(1-\eta)V^{\pm}_{N}+1\Big)
\end{eqnarray}
To find a lower bound on Eve's conditional variances, we first
consider her inference of Bob's state prior to the 50/50 beam
splitter in his station. This is given by
$V_{E|B'}^{\pm}=\min_{g^{\pm}_{E}}
\langle(\hat{X}^{\pm}_{B'}-g'^{\pm}_{E}\hat{X}^{\pm}_{E})^2 \rangle$, where
$'$ labels Bob's state prior to the beam splitter, and $g'^{\pm}_{E}$
is chosen to minimize $V_{E|B'}^{\pm}$.  Eve's measurement variance
after the beam splitter conditioned on Bob's measurement,
$V^{\pm}_{E|B}$, can be expressed in terms of the conditional variance
before the beam splitter, $V^{\pm}_{E|B'}$, as
\begin{eqnarray}\nonumber V^{\pm}_{E|B}&=&{\rm min}_{g^{\pm}_{\rm E}}
\langle(\hat{X}^{\pm}_{B}-g^{\pm}_{E}\hat{X}^{\pm}_{E})^2
\rangle\\
&=& \frac{1}{2}(V^{\pm}_{E|B'}+1)
\end{eqnarray}
where we have used the fact that Eve has no access to the beam
splitter in Bob's station, and therefore has no knowledge of the
vacuum entering through it. The minimum conditional variance
achievable by Alice, prior to the beam splitter in Bob's station,
is $V^{\pm}_{
A|B'(min)}=\eta/V_{A}^{\pm}+(1-\eta)V_{N}^{\pm}$
\cite{G&GReverse}.  Using this fact, and the conditional variance
inequality in Eq.~(\ref{VarUncertainty}), we can establish a lower
bound on Eve's inferences of Bob's measurements
\begin{equation}\label{EvesMinCondVar}
V^{\pm}_{\rm E|B (min)} \geq
\frac{1}{2}\Big(\Big(\frac{\eta}{V^{\pm}_{A}}+(1-\eta)V^{\pm}_{N}\Big)^{-1}+1\Big).
\end{equation}

\begin{figure}[!ht]
\begin{center}
\includegraphics[width=\columnwidth]{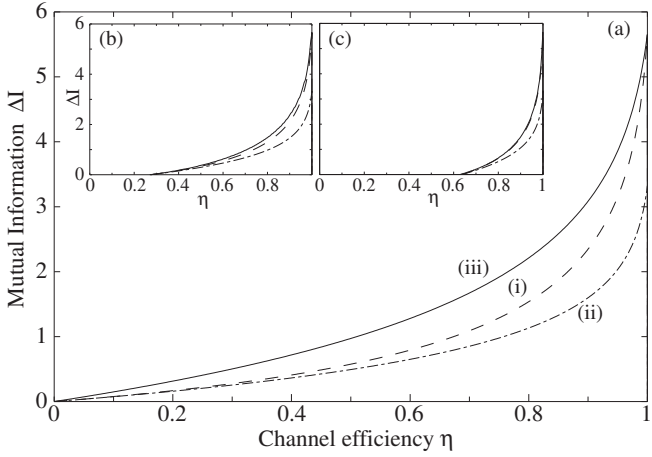}
\caption{Net information rates for the simultaneous and single
quadrature measurement schemes as a function of channel efficiency.
(i) Dashed line, simultaneous quadrature measurement.  (ii) Dot dashed
line, single quadrature measurement.  (iii) Solid line, simultaneous
quadrature measurement with feed forward attack.  For a variance of
$V_{A}=100$ with varying channel noise: (a) V$_{N}$=1 (b) V$_{N}$=1.2
and (c) V$_{N}$=2.}
\label{FigureInformationRates}
\end{center}
\end{figure}

\begin{figure}[!ht]
\begin{center}
\includegraphics[width=\columnwidth]{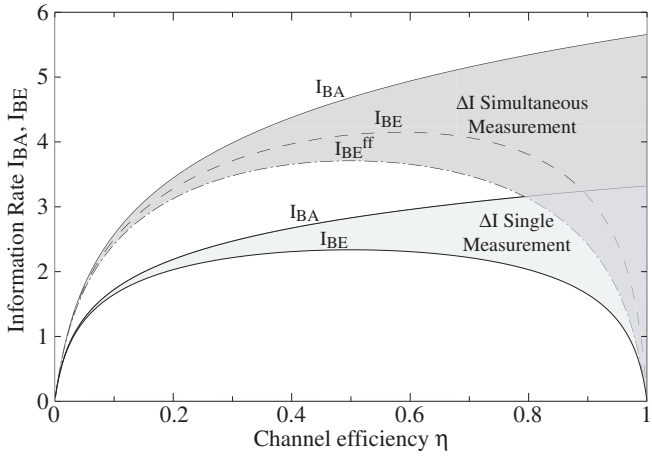}
\caption{Information rates for the simultaneous and single quadrature
measurement schemes as a function of channel efficiency $\eta$, with
$V^{\pm}_{A}\!=\!100$ and $V^{\pm}_{N}\!=\!1$.  The net information
rate for both schemes is ${\Delta I}\!=\!  I_{BA}\!-\!I_{BE}$.  In
the case of simultaneous quadrature measurements, $I_{BE}$, the dashed
line, denotes maximum information rate obtained by Eve, whilst
$I_{BE}^{ ff}$.  the dot dashed line, denotes the information rate
obtained by Eve using the feed forward
attack.}\label{FigureIndividualRates}
\end{center}
\end{figure}

The optimal information rate at which a Gaussian signal can be
transmitted though a channel with additive Gaussian noise is given by
the Shannon formula \cite{Shannon}, which can be expressed as ${
I}=1/2\log_{2} \big(1+S/N\big)$ with units of {\rm (bits/symbol)},
where ${S/N}$ is the standard signal to noise ratio.  This optimal
net information rate can be used to determine the secret key rate for
our simultaneous quadrature measurement protocol, which is given by
$\Delta {I}= \Delta {I}^{+}+\Delta {I}^{-}$, where $\Delta
{I}^{\pm}={I}_{BA}^{\pm}-{I}_{BE}^{\pm}$ with ${
I}_{BA(BE)}$ being the {\it quadrature} information rates between Bob
and Alice (Eve): $I^{\pm}_{BA}\!=\!1/2{\rm
log}_2\big(V^{\pm}_{B}/V^{\pm}_{A|B}\big)$ and
$I^{\pm}_{BE}\!=\!1/2{\rm log}_2\big(V^{\pm}_{B}/V^{\pm}_{E|B}\big)$
\cite{G&GReverse}.  From these expressions, the secret key rate for
the simultaneous quadrature measurement protocol can be expressed as
\begin{eqnarray}\label{DefinitionOfSecretKey}
\Delta {I}&=&\frac{1}{2}\log_{2}
\Big(\frac{V_{E|B}^{+}V_{E|B}^{-}}{V_{A|B}^{+}V_{A|B}^{-}}\Big)
\end{eqnarray}
where the generation of a secret key is only possible when $\Delta
{\rm I}$ is greater than zero. Substituting
Eq.~(\ref{AlicesCondVar}) with $V^{\pm}_{\rm sqz}=1$ (Alice
maximizes her information rate by using coherent states), and
Eq.~(\ref{EvesMinCondVar}) into Eq.~(\ref{DefinitionOfSecretKey}),
gives a lower bound on the secret key rate of
\begin{eqnarray}\label{SecretKeyRateHeterodyne}
\Delta {I}&\geq&\log_{2}
\Big(\frac{(\frac{\eta}{V_{A}}+(1-\eta)V_{N})^{-1}
+1}{\eta+(1-\eta)V_{N}+1} \Big)
\end{eqnarray}
where we have assumed symmetry between the amplitude and phase
quadratures.  Figure~\ref{FigureContour} shows the secret key rate for
the simultaneous quadrature measurement scheme as a function of
channel efficiency and channel noise.  We see that, so long as the
channel noise $V_{N}$ is not excessive, a secret key can be
successfully generated between Alice and Bob, even in the limit of
very small channel efficiency $\eta$.  As the channel noise is
reduced, or efficiency increased, the rate at which a key can be
established is enhanced.  Figure~\ref{FigureInformationRates} compares
the information rates of the simultaneous and single quadrature
protocols as a function of channel efficiency for varying channel
noise.  The information rate for simultaneous quadrature measurements
is always higher than that for single quadrature measurements, and in
the limit of large signal variances and high channel efficiency, it
approaches double.  The individual secret key rates for the
simultaneous and single quadrature measurement protocols can be
calculated and are plotted in Fig.~(\ref{FigureIndividualRates}).  It
should be noted that in our protocol Eve must attempt to determine
Bob's measurements in both the amplitude and phase quadratures.  This
introduces an extra penalty to Eve, which is not included in the lower
bound for her conditional variance in Eq.~(\ref{EvesMinCondVar}).
Therefore, in general, Eve will do even worse than our analysis
suggests.

To establish an upper bound on the secret key rate, we now consider
the physical implementation of an eavesdropping attack for Eve for our
protocol.  In the case where Bob measures a single quadrature,
Grosshans and Grangier showed that an entangling cloner attack
\cite{G&GReverse} is the optimum attack.  However, for simultaneous
quadrature measurements, we found a more effective attack to be a
simple feed forward attack with no entanglement as shown in
Fig.~\ref{FigureSchematic}~(b).  The attack goes as follows: Eve taps
off a fraction of the beam using a beam splitter with transmission
$\epsilon$.  She performs simultaneous quadrature measurements on this
beam, with measured quadrature variances of
\begin{eqnarray}
V_{E}^{\pm}=\frac{1}{2}\Big ((1-\epsilon)V_{
A}^{\pm}+{\epsilon}+1\Big)
\end{eqnarray}
She then applies the measured photocurrents back onto the quantum
channel using electronic feed forward techniques. The variances of
Bob's measurements can then be expressed as
\begin{eqnarray}\label{BobVarforEveAttack}
V_{ B}^{\pm}&=&\frac{1}{2}\Big(\Big(\sqrt{\epsilon}+g_{
E}^{\pm}\sqrt{(1-\epsilon)/2}\Big)^{2} V_{
A}^{\pm}+1\\\nonumber &+&{ V}_\mathcal{N}^{\pm}+g_{
E}^{2\pm}/2+\Big(\sqrt{1-\epsilon}+g_{
E}^{\pm}\sqrt{\epsilon/2}\Big)^{2} \Big)
\end{eqnarray}
where $g_{ E}^{\pm}$ is the gain of the electric feed forward,
and to avoid detection Eve encodes additional Gaussian noise with
a variance ${ V}_\mathcal{N}^{\pm}$ onto the channel.
The gain of Eve's feed forward must be chosen carefully to ensure that
the magnitude of the signal detected by Bob remains invariant.  The
correct gain can be expressed as $g_{E}^{\pm}=
\sqrt{2}(\sqrt\eta-\sqrt\epsilon)/\sqrt{1-\epsilon}$.  Substituting
this into Eq.~(\ref{BobVarforEveAttack}) we obtain Bob's variance due
to the feed forward attack, ${ V_{B}^{ff\pm}}$.  We can now
calculate Eve's conditional variance, $V_{E|B}^{ ff \pm}$, for the
feed forward attack as a function of the beam splitter transmission
$\epsilon$.  Ideally, Eve would take $\epsilon \rightarrow 0$ to gain
as much information about Alice's signal as possible.  However, in
doing so she increases the noise on Alice's inference of Bob's state
and consequently alerts them to her presence.  She must ensure that
her attack does not change the magnitude of this noise.  This places
both lower $\epsilon_{\rm min}$ and upper $\epsilon_{\rm max}$ limits
on the beam splitter transmission.  We numerically minimize
$V_{E|B}^{{\rm ff \pm}}$ for all $\epsilon$ between $\epsilon_{\rm
min}$ and $\epsilon_{\rm max}$, and hence determine the secret key
rate $\Delta { I^{ff}}$.  The secret key rate for the feed forward
attack is plotted in Figs.~(\ref{FigureInformationRates}) and
(\ref{FigureIndividualRates}), and compared with the lower bound
calculated in Eq.~(\ref{SecretKeyRateHeterodyne}).
Figure~(\ref{FigureInformationRates}) shows that for channel noise of
variance $V_{ N}=1$, the feed forward information rate is higher
than our lower bound.  However, as the channel noise variance is
slighty increased, the feed forward bound asymptotes to the lower
bound calculated in Eq.~(\ref{SecretKeyRateHeterodyne}).

To summarize, we have proposed a new coherent state QKD protocol based
on simultaneous quadrature measurements.  We have calculated a lower
bound on the secret key rate for this protocol, finding that in the
limit of large signal variance and high channel efficiency it
approaches twice that of previous coherent state QKD schemes.  We have
considered a possible eavesdropping attack in the form of a simple
feed forward scheme, which has provided us with an upper bound on the
secret key rate.  An important advantage of our simultaneous
quadrature measurement protocol is the increase in total bandwidth.
The absolute information rate, in bits/second, can be expressed as
${ I}={ W} {\rm log}_{2}(1+ S/N)$ \cite{Shannon}, where ${
W}$ is the limiting bandwidth associated with the state preparation or
detection.  Typically, in continuous variable quantum cryptography
schemes, ${\rm W}$ can be attributed to the switching time for the
local oscillator phase.  The simultaneous quadrature measurement
scheme does not require switching, so that orders of magnitude
increases in absolute secret key rates should be achievable.

In conclusion, we have shown that there is no need to randomly switch
bases to achieve secure QKD. By performing simultaneous quadrature
measurements in a coherent state quantum cryptography protocol, we are
able to achieve a significantly larger secret key rate than that
obtained by the usual single quadrature measurements.  This new QKD
protocol will allow simpler and higher bandwidth quantum cryptographic
experiments and technological applications.

We would like to acknowledge the support of the Australian Research
Council and the Australian Department of Defence.  We are grateful to
Roman Schnabel for useful discussions.

\end{document}